\documentclass[journal]{IEEEtran}
\ifCLASSINFOpdf
   \usepackage[pdftex]{graphicx}
\else
\fi
\usepackage{amsmath}
\usepackage{bm}
\ifCLASSOPTIONcompsoc
  \usepackage[caption=false,font=normalsize,labelfont=sf,textfont=sf]{subfig}
\else
  \usepackage[caption=false,font=footnotesize]{subfig}
\fi
\begin{document}
%
\title{Principal Cross-Border Flow Patterns in the European Electricity Markets}
\author{\IEEEauthorblockN{
Mirko Sch{\"a}fer\IEEEauthorrefmark{1},
Fabian Hofmann\IEEEauthorrefmark{2},
Hazem Abdel-Khalek\IEEEauthorrefmark{1} and
Anke Weidlich\IEEEauthorrefmark{1}}\\
\IEEEauthorblockA{\IEEEauthorrefmark{1}Department of Sustainable Systems Engineering (INATECH), University of Freiburg, 79110 Freiburg, Germany\\
Email: mirko.schaefer@inatech.uni-freiburg.de}\\
\IEEEauthorblockA{\IEEEauthorrefmark{2}Frankfurt Institute for Advanced Studies, 60438~Frankfurt~am~Main, Germany}%

\thanks{
{\copyright} 2019 IEEE. Personal use of this material is permitted. Permission from IEEE must be obtained for all other uses, in any current or future media, including reprinting/republishing this material for advertising or promotional purposes, creating new collective works, for resale or redistribution to servers or lists, or reuse of any copyrighted component of this work in other works.}
}
\IEEEoverridecommandlockouts

%

\maketitle

\begin{abstract}
The interconnected European Electricity Markets see considerable cross-border trade between different countries. In conjunction with the structure and technical characteristics of the power grid and its operating rules, the corresponding commercial flows translate into actual physical flows on the interconnection lines. From the interplay of different physical, technical, and economic factors thus emerge complex spatiotemporal power flow patterns.
Using Principal Component Analysis, in this contribution hourly time-series of cross-border physical flows between European countries in 2017 and 2018 are analyzed. The most important patterns in the time series of imports/exports and cross-border physical flows are identified. Their spatial and temporal structure, as well as their contribution to the overall variance is described. Additionally, we apply a tracing technique to the overall flow patterns, which allows identifying the physical power transfers between European countries through the common grid infrastructure.
\end{abstract}


\IEEEpeerreviewmaketitle
\section{Introduction}
The increasing integration of European electricity markets is considered to be an essential factor for the European Unions aim to ensure secure, affordable and sustainable energy supply for its citizens~\cite{energy-union-package}. This integration is advanced both by the expansion of power transmission capacities~\cite{tyndp2018} as well as policy measures working towards more efficient use of the existing infrastructure, for instance by adopting more elaborated market coupling mechanisms~\cite{vandenbergh2016}. It has been reported that this process already has made significant progress, with energy now being traded more freely across borders~\cite{report_energy-union2019}. These trades are reflected in considerable physical power flows between the European countries, adding up to approximately $435$ TWh of overall cross-border exchanges of ENTSO-E countries in 2018~\cite{fact_sheet_2018}. This corresponds to $8.4\%$ of the aggregated electricity consumption in the ENTSO-E area~\cite{fact_sheet_2018}. Only in recent years, detailed data with hourly resolution describing these cross-border flows have been made publicly available through the ENTSO-E Transparency Page~\cite{transparency, hirth2018}. Ref.~\cite{tranberg2019} uses this cross-border flow data to implement a real-time consumption-based carbon accounting approach based on the application of the flow tracing technique~\cite{hoersch2018}. Such an approach is also discussed by the electricityMap project, which visualizes cross-border flows and associated CO$_2$ emissions~\cite{electricitymap}. Applying a more straightforward analysis of the data, the recent market report from the German TSO Amprion analyses selected physical flows for the time between May 2015 and November 2018~\cite{amprion2019}. Nevertheless, to our knowledge a detailed analysis or well-documented usage of this data up to now is mostly missing in the literature.

In this contribution, we present an analysis of hourly cross-border flow data from 2017-2018 with the aim to identify patterns in this high-dimensional data. The cross-border physical flows are the emergent result of a multitude of factors including the spatial distribution and temporal evolution of generation and load in the countries, the trades realized between market participants from different countries, and the topology and operation of the grid infrastructure. In order to identify patterns in the resulting fluctuation flows, we apply Principal Component Analysis (PCA) to the time series of countries net import/export and cross-border physical flows on most borders in the ENTSO-E area. The general idea of PCA is to reduce the complexity of a multivariate data set of dimension $N$ by choosing a reduced set of $M<N$ axes, which still represent most of the variance of the original system. The new axes correspond to patterns associated with decreasing variance in the data set. Recently, PCA has been applied to detect mismatch, nodal injection, and flow patterns in models of a highly renewable European electricity system~\cite{raunbak2017, hofmann2018}. An analysis similar to the approach taken in this contribution is presented in a recent Amprion market report, where PCA is applied to commercial flows in Central Western Europe~\cite{amprion2019}. While the PCA analysis targets patterns in the fluctuations of net imports/exports and cross-border flows, the average flows are analyzed using the flow tracing technique. This method was originally proposed in the context of loss allocation in electricity grids~\cite{kirschen1997, bialek1996}, but has recently been applied to analyze transmission usage, flow-based patterns of import and export, and storage usage in energy system models~\cite{tranberg2015, hoersch2018, tranberg2018}. It allows to follow electricity in the power grid from the sources to the sinks, and thus providing an intuitive measure of physical import/export patterns.

The article is structured as follows: Section~\ref{sec:data} describes the data sources and processing, while Section~\ref{sec:methods} briefly reviews Principal Component Analysis and the flow tracing method. In Section~\ref{sec:results} results are presented, and Section~\ref{sec:conclusion} draws conclusions and gives an outlook to possible extensions of this study.
%
%
\section{Data}
\label{sec:data}
The European Network of Transmission System Operators for Electricity (ENTSO-E) publishes data on cross-border physical flows both through its Power Statistics web page~\cite{power_statistics} and the ENTSO-E Transparency Platform ~\cite{transparency}. The Power Statistics web page provides consolidated data, which takes into account national statistical resources~\cite{monthly_statistics}. Its aggregated values correspond to the yearly flows as published in the official ENTSO-E statistical fact sheet~\cite{fact_sheet_2018}. Whereas this data set thus can be considered to be reliable and without gaps, it only gives monthly aggregated and additionally selected values and therefore does not provide the temporal resolution needed for a more detailed analysis of physical flows between European countries. In contrast, the ENTSO-E Transparency Platform provides this data with an hourly resolution. Although it has been reported that the quality of many data items published on this platform has recently been improved~\cite{hirth2018}, in particular the hourly time series for cross border physical flows show gaps in the data for several borders, with the aggregated values often not corresponding to the monthly values published on the Power Statistics web page. Using both these sources we construct a data set describing physical flows with an hourly resolution for a selected set of borders between countries in the ENTSO-E area by performing the following steps.

The data from the Transparency Platform for cross-border physical flows for 2017 and 2018 is downloaded using the File Protocol Server (we exclude earlier data due to an increasing number of gaps, and later data because it might still be corrected by ENTSO-E). We then determine the number $g_{l}(\tau)$ of gaps with length $\tau$ for each border $l$ and calculate the measure
\begin{align}
    G_{l}^{\alpha}=\sum_{\tau=1}^{\tau_{\max}}g(\tau)\tau^{\alpha}~.   
    \label{eq:threshold}
\end{align}
For $\alpha=1$ this measures counts the number of missing hourly values, whereas a parameter $\alpha>1$ gives higher weight to longer gaps in the data. Choosing the parameter $\alpha=1.5$ we find that $58$ out of $78$ borders have a value $G_{l}^{\alpha}<200$, with $36$ borders being represented by a complete data sets with $G_{l}^{\alpha}=0$, and the data sets for $6$ borders only missing one hour, that is $G_{l}^{\alpha}=1$. The border with the largest value $G_{l}^{\alpha}\approx 190$ in the range $[0,200]$ is Greece-Italy, with this value mainly determined by one gap of $24$ hours, and one gap of $13$ hours. After removing the data for the $20$ borders with $G_{l}^{\alpha}>200$, the total value of represented absolute flows corresponds to approximately 88\% of the total absolute flow on the entire network published on the ENTSO-E Power Statistics web page.

\begin{figure}[t!]
\centering
\includegraphics[width=0.95\linewidth]{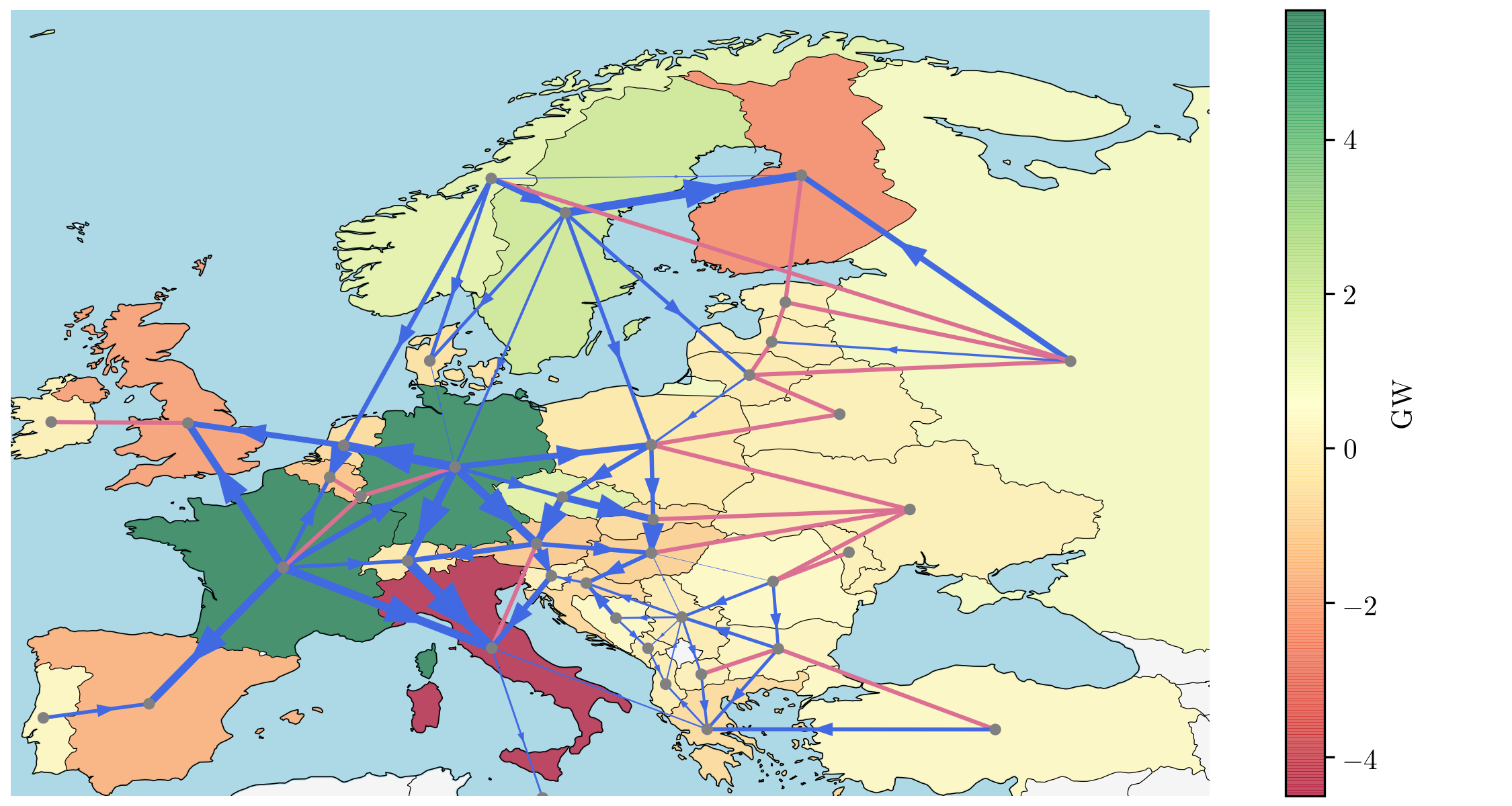}
\caption{Average net imports/exports $\langle p_{n}(t)\rangle$ and physical cross-border flows $\langle f_{l}(t)\rangle$ for the selected borders between ENTSO-E countries.}
\label{fig:avg_flow_injection}
\end{figure}

The gaps in this filtered data set are filled using an average week for each border. For each month and border, the hourly time series then is scaled such that the aggregate corresponds to the monthly value published on the Power Statistics web page. From the resulting time series of hourly flows $f_{l}(t)$ for the $58$ selected borders, we calculate for each hour the nodal injection $p_{n}(t)$ of the $33$ included countries:
\begin{align}
    p_{n}=\sum_{l}K_{nl}f_{l}(t)~.
\label{eq:def_nodal}
\end{align}
Here $K_{ln}$ denotes the networks incidence matrix,
\begin{align}
    K_{nl}=\begin{cases}
+1, & \text{link } l \text{ starts at node }n\\
-1, & \text{link } l \text{ ends at node }n\\
0, & \text{else}
\end{cases}~.
\end{align}
Fig.~\ref{fig:avg_flow_injection} shows the network represented by our final data set. Pink coloured links indicate borders with monthly cross-border values published in the ENTSO-E Power Statistics, but which have been removed due to the hourly values from the Transparency Platform failing the criteria in Eq.~(\ref{eq:threshold}). It is apparent that mainly physical cross-border flows in the East of Europe are not yet covered reliably in the data published on the Transparency Platform, but also for instance the data for flows associated with Luxembourg and between Ireland and the UK shows considerable gaps. The countries in Fig.~\ref{fig:avg_flow_injection} are colour-coded according to their average net import/export $\langle p_{n}(t)\rangle$, calculated according to Eq.~(\ref{eq:def_nodal}) from the cross-border physical flows represented in our data set. Note that due to the borders removed from the data set, the aggregated nodal injections, in general, do not correspond to the aggregated balances as published in the ENTSO-E statistical factsheet~\cite{fact_sheet_2018}. Nevertheless, France and Germany as the largest net exporters, and Italy as the largest net importer are clearly indicated in our data set.

%

%
%
\section{Methods}
\label{sec:methods}
%
%
%
\begin{figure}[!t]
\centering
\includegraphics[width=\linewidth]{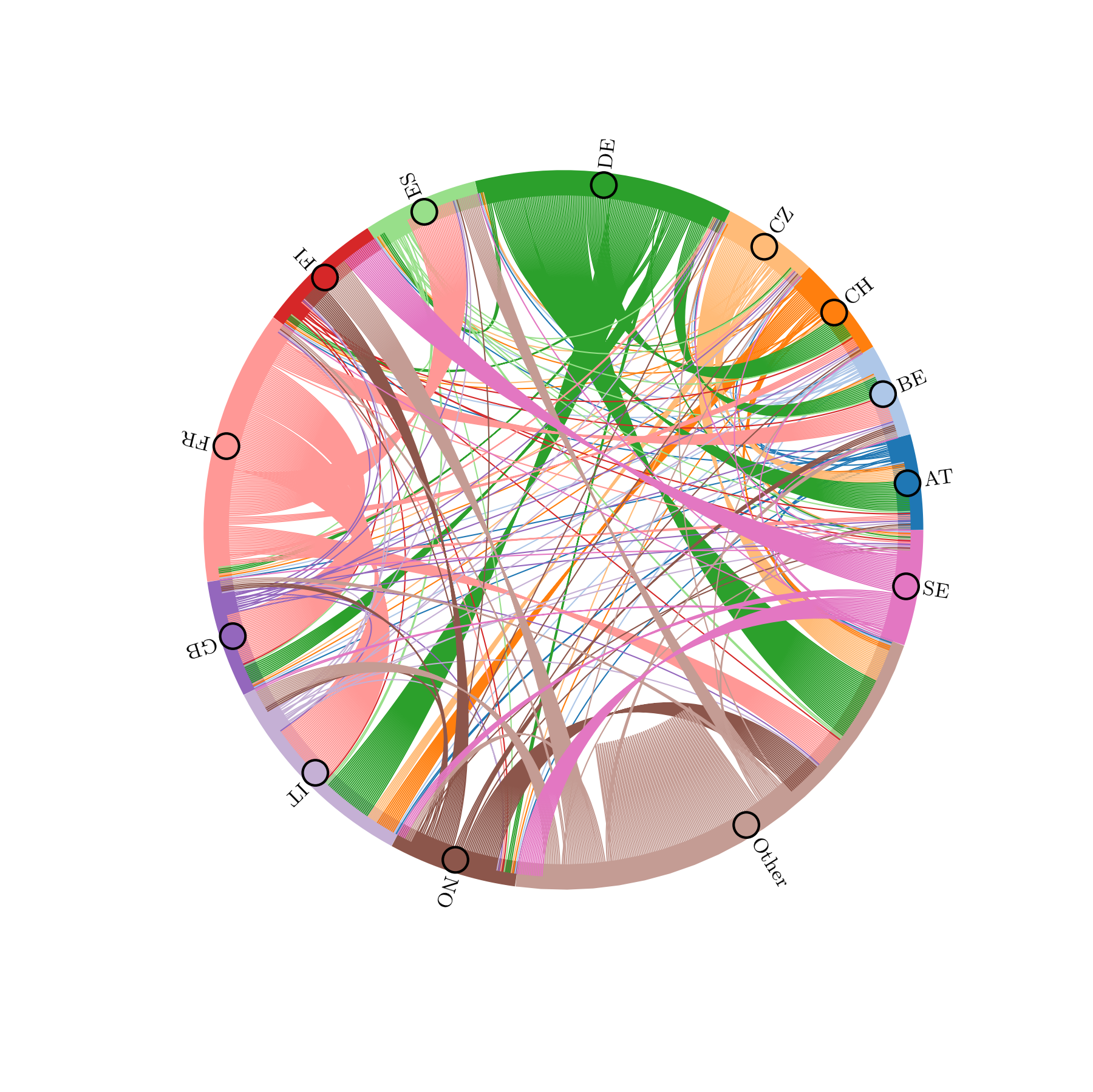}
\caption{Average power transfer $\langle T_{m}^{n}(t)\rangle$ from country $m$ to country $n$ derived by means of the flow tracing algorithm applied to the hourly physical cross-border flows $f_{l}(t)$.}
\label{fig:flowtracing}
\end{figure}
%
The \emph{flow tracing} algorithm is applied to each flow pattern $f_l(t)$ and associated net import/export pattern $p_{n}(t)$. We define the net nodal export as $p_{n}^{\mathrm{ex}}(t)=\max[p_{n}(t),0]$, and the net nodal import as $p_{n}^{\mathrm{im}}(t)=\max[-p_{n}(t),0]$. Flow conservation at every time step $t$ is expressed by the equation
\begin{align}
    p_{n}^{\mathrm{ex}}(t) + \sum_{k}f_{k\to n}(t) = p_{n}^{\mathrm{im}}(t) + \sum_{k} f_{n\to k}(t)~,
\end{align}
where $f_{n\to k}(t)$ denotes the flow from node $n$ to node $k$ at time $t$. The method of flow tracing follows the different nodal exports $p_{n}^{\mathrm{ex}}$ downstream through the network. Nodal inflows of different origin are assumed to perfectly mix inside the node, thus determining the composition of outgoing flows and nodal imports~\cite{tranberg2015}. This approach allows to calculate the share $q_{n,m}(t)$ of the nodal import $p_{n}^{\mathrm{im}}$ at node $n$ which is associated with the export $p_{m}^{\mathrm{ex}}$ from node $m$ using the following equation for partial flow conservation:
\begin{align}
    \delta_{n,m}p_{n}^{\mathrm{ex}}+\sum_{k}  q_{k,m}f_{k\to n}=
     q_{n,m}p_{n}^{\mathrm{im}} + \sum_{k}q_{n,m}f_{n\to k}~.
\label{eq:partial_flows}
\end{align}
For simplicity here the time index $t$ has been omitted. Equation~(\ref{eq:partial_flows}) can be rewritten as the following matrix equation:
\begin{align}
    \delta_{n,m}p_{n}^{\mathrm{ex}}=
    \sum_{k}\left[
    \delta_{n,k}\left(
    p_{n}^{\mathrm{im}}+\sum_{k'}f_{n\to k'}\right)
    -f_{k\to n}\right]
    q_{k,m}~.
\end{align}
This matrix equation can be inverted, yielding the share of imports $q_{n,m}(t)$ in country $n$ associated with an export from country $m$, calculated from the set of the nodal imports/exports $p_{n}(t)$ and flows $f_{l}(t)$~\cite{hoersch2018}. The associated value $T^{n}_{m}=q_{n,m}(t)p_{n}^{\mathrm{im}}$ describes the country-to-country transfer, that is the amount of flow exported from country $m$ through the power grid to the importing country $n$ at time $t$.

\emph{Principal Component Analysis (PCA)} is applied to both the time series of cross-border physical flows $f_{l}(t)$ and the associated net imports/exports $p_{n}(t)$. Note that although flows and nodal injections are related through Eq.~(\ref{eq:def_nodal}), the connection between their principal components in general is non-trivial~\cite{hofmann2018}. The principal import/export components $\bm{\rho}_{k}^{p}$ are calculated as the normalized eigenvectors of the covariance matrix $\bm{\Sigma}^{p}=\text{cov}(\bm{p},\bm{p})$. The associated normalized eigenvalues $\tilde{\lambda}_{k}^{p}$ of $\bm{\Sigma}^{p}$,
\begin{align}
    \tilde{\lambda}_{k}^{p}=\frac{\lambda_{k}^{p}}{\sum_{k'}\lambda_{k'}^{p}}=
    \frac{\lambda_{k}^{p}}{\text{tr}\:(\bm{\Sigma}^{p})}~,
\end{align}
are a measure of the share of the total variance of the data in the direction of the new axis $\bm{\rho}^{p}_{k}$. Due to the normalization it holds $\sum_{k}\tilde{\lambda}^{p}_{k}=1$. The principal flow components $\bm{\rho}^{f}_{k}$ with associated normalized eigenvalues $\tilde{\lambda}_{k}^{f}$ are calculated analogously based on the flow covariance matrix $\bm{\Sigma}^{f}=\text{cov}(\bm{f},\bm{f})$. The time evolution of the components is given by the amplitudes $\beta_{k}^{p}(t)$ and $\beta_{k}^{f}(t)$, which are calculated by projecting the corresponding mean-free data onto the axis expressing the specific components:
\begin{align}
    \bm{p}(t)-\langle\bm{p}(t)\rangle=\sum_{k}\beta_{k}^{p}(t)\bm{\rho}_{k}^{p}~,\nonumber\\
    \bm{f}(t)-\langle\bm{f}(t)\rangle=\sum_{k}\beta_{k}^{f}(t)\bm{\rho}_{k}^{f}~.
\label{eq:amplitudes}
\end{align}
\section{Results}
\label{sec:results}

%
\begin{figure}[!t]
\centering
\includegraphics[width=\linewidth]{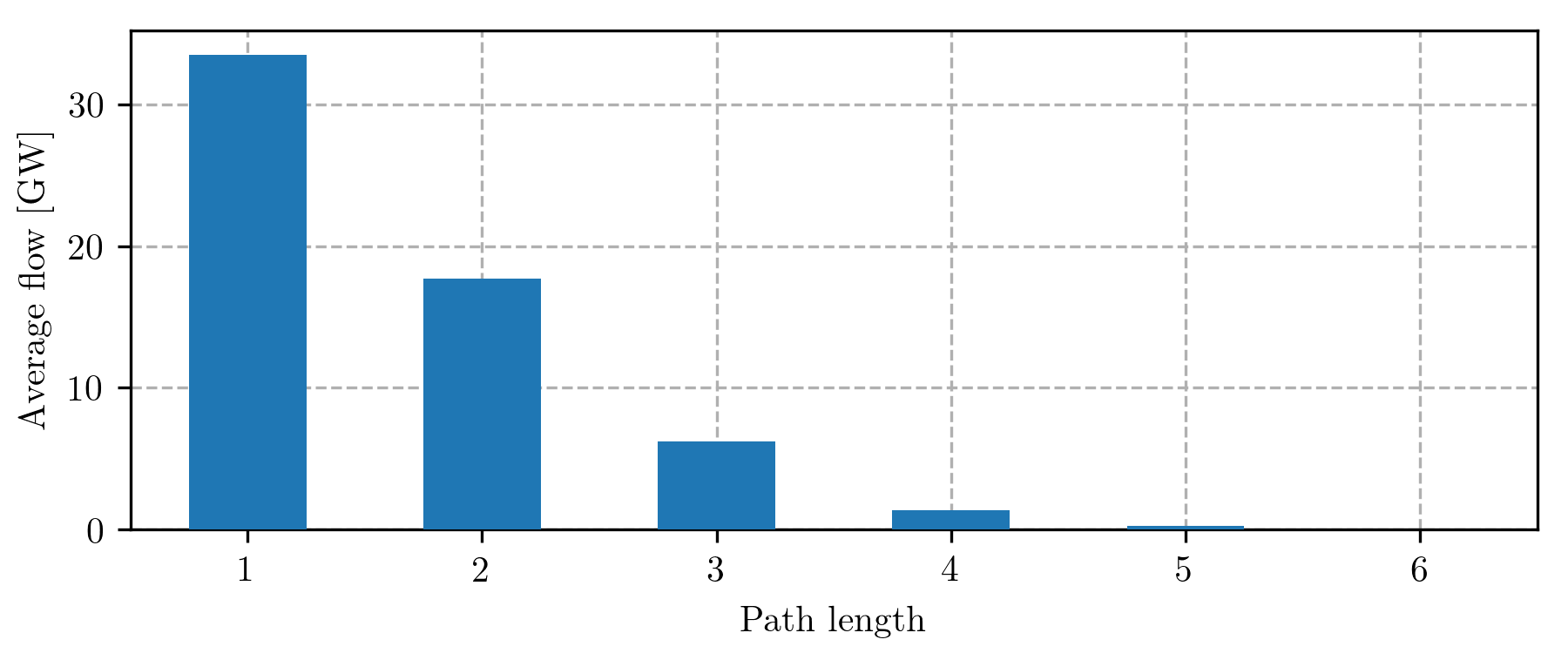}
\caption{Aggregated average transfers $\langle T_{m}^{n}(t)\rangle$ between countries $m$ and $n$ according to the number of crossed country borders on the shortest path between them (path length).}
\label{fig:distance}
\end{figure}
%
Fig.~\ref{fig:flowtracing} visualizes the average transfers $\langle{T}_{m}^{n}(t)\rangle$ between selected countries based on the application of the flow tracing technique to the hourly cross-border flows $f_{l}(t)$. Only the ten countries with the largest average net import plus net export are shown explicitly, the remaining countries are aggregated as `Other'. Line segments on the circle are scaled according to the sum of average exports and average imports. The links are scaled according to the corresponding average transfer $\langle T_{m}^{n}(t)\rangle$. Since this algorithm is applied to each hour separately, in general one observes transfers in both directions between all considered pairs of countries.

The visualization in Fig.~\ref{fig:flowtracing} emphasizes the role of large cross-border flows between neighboring countries, for instance between France and Germany, or between France and Spain. Also considerable transfers over two borders can be identified, prominently between Germany and Italy or between Germany and Great Britain. These relations are also suggested from the average flow pattern displayed in Fig.~\ref{fig:avg_flow_injection}. Nevertheless, Fig.~\ref{fig:flowtracing} illustrates the interconnectedness of the  system by showing non-vanishing transfers between almost all countries, with these transfers often crossing several borders. This is displayed explicitly in Fig.~\ref{fig:distance}, which shows the average physical transfer between countries dependent on the number of borders crossed on the shortest path between them (path length). One observes that a considerable share of transfers happens between countries that are separated by more than two borders. This finding emphasizes the importance of transmission infrastructure beyond the adjacent borders for each country~\cite{tyndp2018}, and the necessity of compensation mechanisms taking into account transient flows~\cite{itc, acer2013}.

%
\begin{figure*}[!t]
\centering
\subfloat{\includegraphics[width=0.95\linewidth]{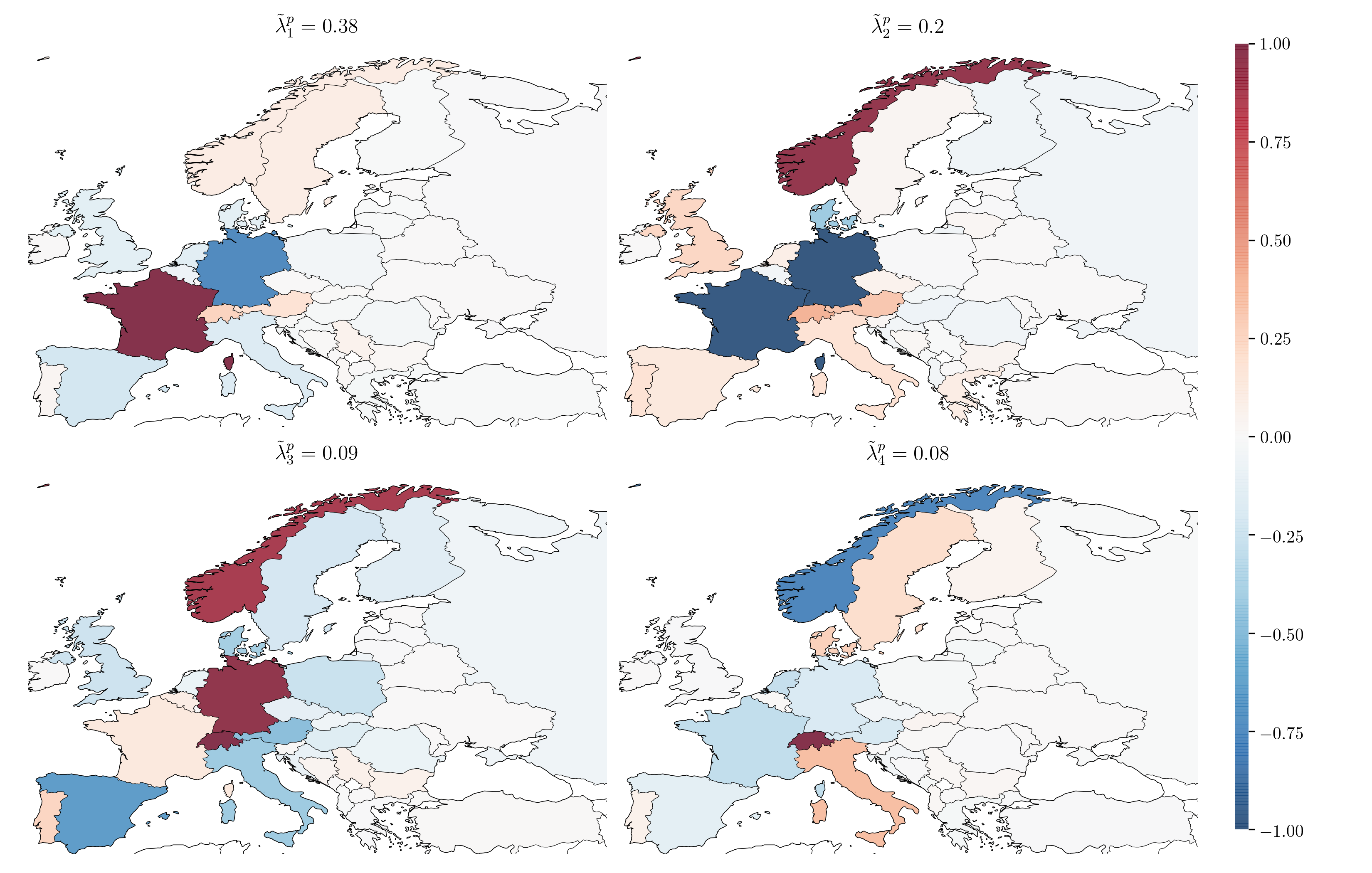}
}
\caption{The first four principal import/export patterns $\bm{\rho}_{k}^{p}$ with associated normalized eigenvalue $\tilde{\lambda}_{k}^{p}$.}
\label{fig:nodal_components}
\end{figure*}
\begin{figure}[!t]
\centering
\includegraphics[width=\linewidth]{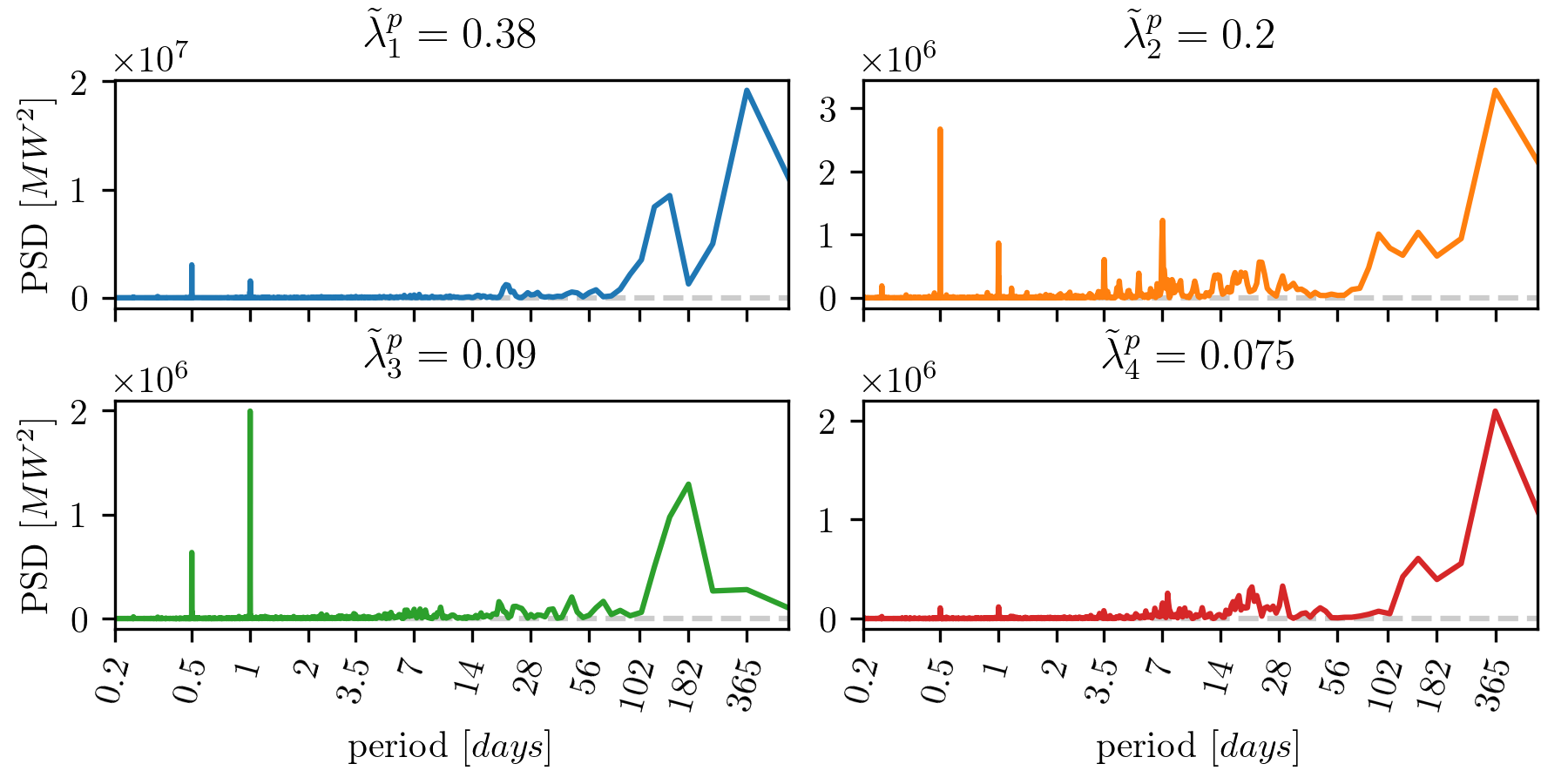}
\caption{Power Spectral Densities (PSD) of the amplitudes $\beta_{k}^{p}(t)$ describing the time evolution of the first four principal import/export components.}
\label{fig:fourier_injection}
\end{figure}

%
Results from the application of PCA to the time series of net imports/exports $p_{n}(t)$ are visualized in Fig.~\ref{fig:nodal_components}. This figure shows the first four principal import/export components. The first component describes a dipole between France and Germany, which is an anti-correlated import/export pattern of these countries. Remarkably, this pattern already can be associated with $\tilde{\lambda}_{1}^{p}=38\%$ of the variation in the import/export time-series. The second component, associated with $\tilde{\lambda}_{2}^{p}=20\%$ of the variation, depicts a dipole between France and Germany on the one hand, and Norway on the other hand. The third component with $\tilde{\lambda}_{3}^{p}=9\%$ shows a strong correlation between the import/export of Germany and Norway, whereas the fourth component ($\tilde{\lambda}_{4}^{p}=8\%$) can be associated with an anti-correlation of the import/export of Norway and Italy/Switzerland. The first four components together already cover approximately $75\%$ of the variance in the data, with the other $29$ components combined covering the remaining $25\%$ of the variance. The temporal evolution of these patterns is defined by the amplitudes $\beta_{k}^{p}(t)$, see Eq.~(\ref{eq:amplitudes}). In Fig.~\ref{fig:fourier_injection} the Fourier transform of the first four amplitudes $\beta_{k}^{p}(t)$ is described by the corresponding Power Spectral Densities (PSD). It can be observed that the first and fourth components are mainly connected to seasonal cycles, whereas the second and third components also have a strong relationship to diurnal and half-diurnal cycles.

%
\begin{figure*}[!t]
\centering
\subfloat{\includegraphics[width=0.95\linewidth]{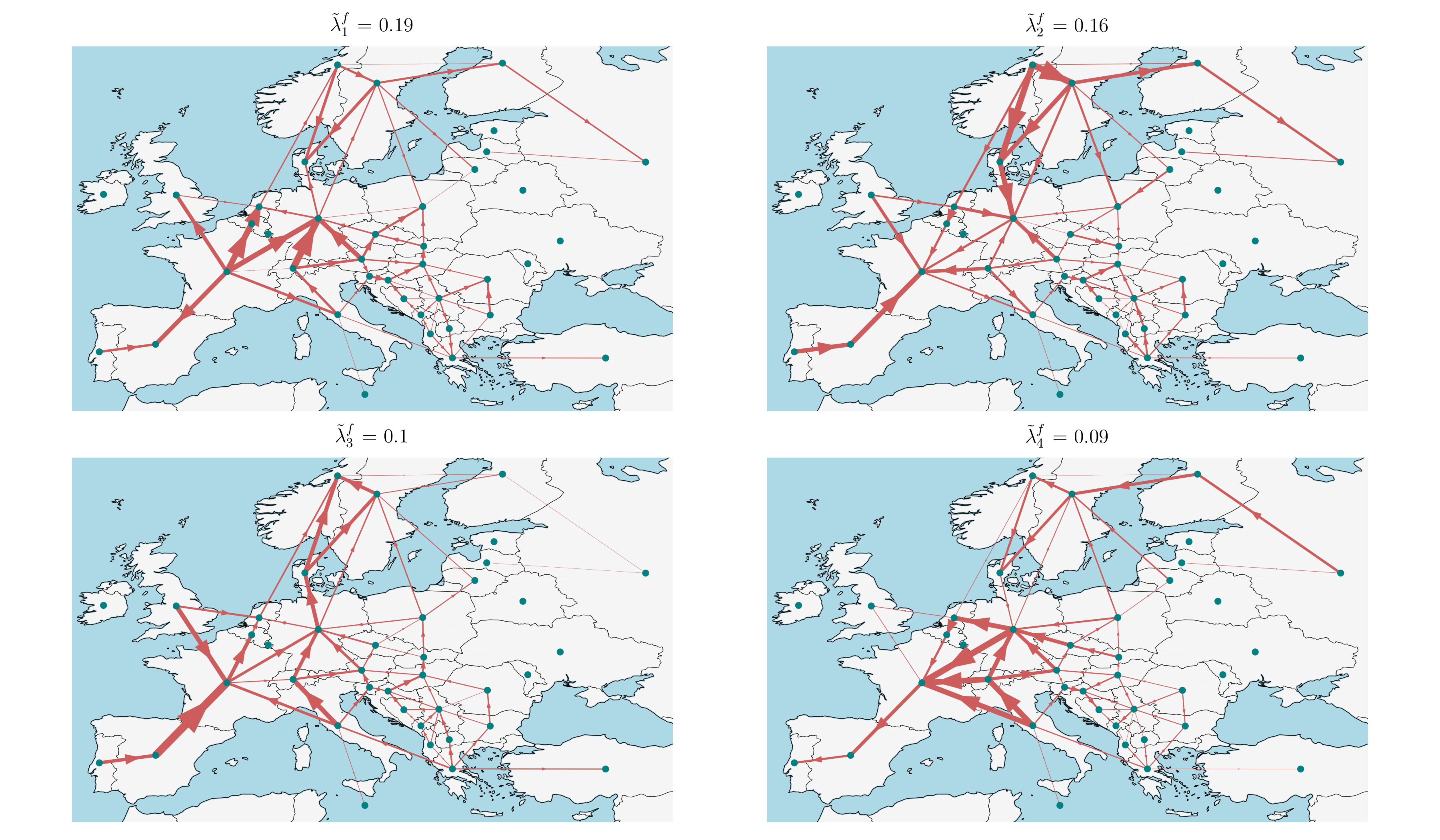}
}
\caption{The first four principal flow patterns $\bm{\rho}_{k}^{f}$ with associated normalized eigenvalue $\tilde{\lambda}_{k}^{f}$.}
\label{fig:flow_components}
\end{figure*}

\begin{figure}[!t]
\centering
\includegraphics[width=\linewidth]{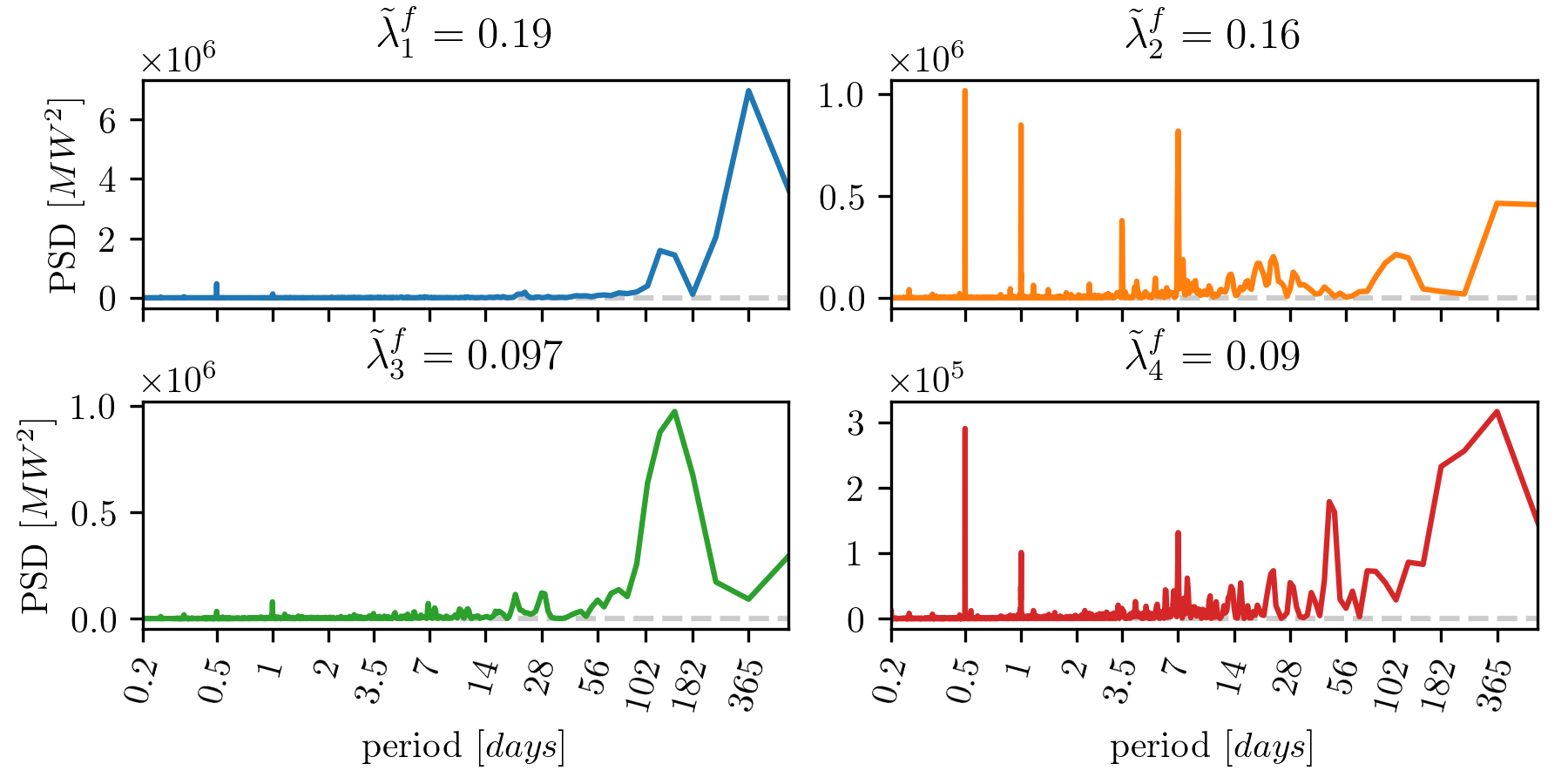}
\caption{Power Spectral Densities (PSD) of the amplitudes $\beta_{k}^{f}(t)$ describing the time evolution of the first four principal flow components.}
\label{fig:fourier_flow}
\end{figure}

%
Fig.~\ref{fig:flow_components} displays the first four principal components of the cross-border physical flow time series $f_{l}(t)$. The net imports/exports corresponding to these patterns, in general, are different from the principal components displayed in Fig.~\ref{fig:nodal_components}~\cite{hofmann2018}. The first flow component ($\tilde{\lambda}_{1}^{f}=19\%$) shows strong outgoing flows from France, and strong ingoing flows into Germany. In contrast, the second component ($\tilde{\lambda}_{2}^{f}=16\%$) highlights outgoing flows from Norway through Sweden and Denmark towards Germany, and flows from the Iberian Peninsula towards France. The latter flows are even more pronounced in the third component ($\tilde{\lambda}_{3}^{f}=10\%$), and occur in combination with flows towards the Scandinavian countries. The fourth component ($\tilde{\lambda}_{4}^{f}=9\%$) shows strong outgoing flows from Germany and Italy.  
%
%
The associated PSD are displayed in Fig.~\ref{fig:fourier_flow}. One observes seasonal cycles for all of these patterns. The second and fourth patterns also show strong half-diurnal and diurnal cycles, with the second component further being related to cycles of half a week and a week.
%
%
%
\section{Conclusion}
\label{sec:conclusion}
This contribution presents the application of  Principal Component Analysis (PCA) and flow tracing to a data set describing hourly cross-border physical flows between European countries for the years 2017 and 2018. By tracing power flows from the exporting countries through the network to the importing countries, their physical transfer of power flows can be estimated. It is shown that although the majority of transfers happen between neighboring countries, a considerable share of electricity trade leads to physical transfers crossing more than one border. The application of PCA shows that for both the import/export and cross-border flow time series, the first four principal components represent a significant share of the total variance ($75\%$ for the imports/exports, $54\%$ for the flows). The main import/export patterns highlight correlations or anti-correlations between specific countries, for instance, anti-correlation of imports/exports of France and Germany in the first principal pattern, or correlation between France and Germany, and anti-correlation with Norway in the second principal pattern. The associated time evolution shows strong seasonal, and depending on the pattern, also (half-) diurnal or weekly cycles. The first principal flow patterns are composed of strong flows either to or from Germany, France, the Scandinavian countries, and the Iberian Peninsula. Also for these patterns, the time evolution shows seasonal cycles, and for some patterns also (half-) diurnal and (half-) weekly patterns.

The presented analysis has identified and described specific spatiotemporal patterns in the network of cross-border physical flows emerging from the interconnected European electricity markets. These findings provide a first step towards an understanding of the interplay between generation and load patterns, market dynamics, and the role of the physical constraints imposed by the transmission infrastructure. For a deeper interpretation of the observed patterns, it would be necessary to consider more detailed data describing the generation patterns inside the countries, for instance, the deployment of renewable energy sources in Germany. Further insights could be obtained by comparing the cross-border physical flow patterns to the corresponding commercial flow patterns. It would also be interesting to study how the principal patterns evolve over time, and how this evolution is affected by changes in the market operation or the underlying transmission infrastructure.

%
%
\ifCLASSOPTIONcaptionsoff
  \newpage
\fi



%
\bibliographystyle{IEEEtran}
\bibliography{references}

\end{document}